%% file: main.tex
\documentclass[twocolumn]{aastex63}

\usepackage{rotating}
\usepackage{acronym}
\usepackage{xspace}
\usepackage{bbm}
\usepackage{amsmath}
\usepackage{mathtools}
\usepackage{graphicx}

\graphicspath{{./}{figures/}}

\input{commands}

\input{values}
\input{acronyms}

\definecolor{chmagenta}{rgb}{0.54, 0.17, 0.88}

\definecolor{tablegreen}{rgb}{0.0, 0.5, 0.0}
\definecolor{tableorange}{rgb}{0.86, 0.52, 0.0}
\definecolor{tablered}{rgb}{0.7, 0.13, 0.13}

\newcommand{\CIERA}{\affiliation{Center for Interdisciplinary Exploration and Research in Astrophysics (CIERA) and Department of Physics and Astronomy, Northwestern University, 1800 Sherman Ave, Evanston, IL 60201, USA}}
\newcommand{\PADOVAUNI}{\affiliation{Physics and Astronomy Department Galileo Galilei, University of Padova, Vicolo dell'Osservatorio 3, I-35122 Padova, Italy}}
\newcommand{\PADOVAVIRGO}{\affiliation{INFN - Padova, Via Marzolo 8, I--35131 Padova, Italy}}

\shorttitle{Exploring the Lower Mass Gap}
\shortauthors{Zevin, Spera, Berry, \& Kalogera}

\begin{document}

\title{Exploring the Lower Mass Gap and Unequal Mass Regime in Compact Binary Evolution}

\author[0000-0002-0147-0835]{Michael Zevin}\thanks{zevin@u.northwestern.edu}
\CIERA

\author[0000-0003-0930-6930]{Mario Spera}
\CIERA
\PADOVAUNI
\PADOVAVIRGO

\author[0000-0003-3870-7215]{Christopher P L Berry}
\CIERA
\affiliation{SUPA, School of Physics and Astronomy, University of Glasgow, Glasgow G12 8QQ, UK}

\author[0000-0001-9236-5469]{Vicky Kalogera}
\CIERA

\begin{abstract}

On 2019 August 14, the LIGO and Virgo detectors observed GW190814, a gravitational-wave signal originating from the merger of a $\simeq 23 M_\odot$ black hole with a $\simeq 2.6 M_\odot$ compact object.
GW190814's compact-binary source is atypical both in its highly asymmetric masses and in its lower-mass component lying between the heaviest known neutron star and lightest known black hole in a compact-object binary. 
If formed through isolated binary evolution, the mass of the secondary is indicative of its mass at birth. 
We examine the formation of such systems through isolated binary evolution across a suite of assumptions encapsulating many physical uncertainties in massive-star binary evolution. 
We update how mass loss is implemented for the neutronization process during the collapse of the proto-compact object to eliminate artificial gaps in the mass spectrum at the transition between neutron stars and black holes. 
We find it challenging for population modeling to match the empirical rate of GW190814-like systems whilst simultaneously being consistent with the rates of other compact binary populations inferred from gravitational-wave observations.
Nonetheless, the formation of GW190814-like systems at any measurable rate requires a supernova engine model that acts on longer timescales such that the proto-compact object can undergo substantial accretion immediately prior to explosion, hinting that if GW190814 is the result of massive-star binary evolution, the mass gap between neutron stars and black holes may be narrower or nonexistent. 

\end{abstract}


\section{Introduction}\label{sec:intro}

The third observing run of the Advanced LIGO--Virgo network~\citep{aLIGO,aVirgo} has already yielded unprecedented discoveries: the most massive \ac{BNS} system~\citep{GW190425}, and compact binaries with significantly asymmetric masses~\citep{GW190412, GW190814}. 
The most recently announced event, GW190814~\citep{GW190814}, also had a compact-object component that lies within the observed gap in masses between \acp{NS} and \acp{BH}, known as the \acl{LMG}~\citep[\acs{LMG};][]{Bailyn1998, Ozel2010, Farr2011, Ozel2012}. 
Since no tidal signatures are measurable in the \ac{GW} data and no electromagnetic or neutrino counterpart has been reported~\citep[][and references therein]{GW190814}, the nature of the lighter object in the binary is uncertain. 
Nonetheless, this event establishes that \SecondaryMassApprox compact objects do exist in binaries. 

The majority of non-recycled \acp{NS} in the Galaxy have masses of ${\sim\,1.3\,\Msun}$~\citep{Ozel2012,Kiziltan2013}. 
However, the maximum mass that an \ac{NS} can achieve, $m_\mathrm{NS}^\mathrm{max}$, is currently uncertain. 
The Galactic millisecond pulsar J0740+6620 has a Shapiro-delay mass measurement of \CromatieMSP~\citep[$68\%$ credibility;][]{Cromartie2020}; this has been updated to \CromatieMSPUpdated when analyzed using a population-informed prior~\citep{Farr2020}. 
The pulsar J1748$-$2021B has been estimated to have a mass of \FriereMSP ($68\%$ confidence) assuming that the periastron precession is purely relativistic, but if there are contributions from the tidal or rotational distortion of the companion, this estimate would not be valid~\citep{Freire2008a}. 
GW190425's primary component had a mass greater than most Galactic \acp{NS}, \MassRatioNinetyAprilGW~\citep[$90\%$ credibility;][]{GW190425}; 
while high \ac{NS} masses of $\lesssim\,2.5\,\Msun$ can be explained theoretically via stable accretion in low- and intermediate-mass \ac{XRB} systems~\citep[e.g.,][]{Pfahl2003, Lin2011, Tauris2011}, the explanation for a high-mass component in a \ac{BNS} system is open to debate~\citep[e.g.,][]{Romero-Shaw2020,Safarzadeh2020d,Kruckow2020}. 
These observational constraints provide key insights about the \ac{NS} \ac{EOS}; although candidate \acp{EOS} for non-rotating \acp{NS} can have maximum masses that extend as high as $\sim 3\,\Msun$~\citep{Rhoades1974,Muller1996,Kalogera1996a}, population studies of known \acp{NS}~\citep{Alsing2018, Farr2020}, analysis of the tidal deformability of GW170817~\citep{GW170817_EoS,Essick2020}, modeling of the electromagnetic counterparts associated with GW170817~\citep{GW170817_MMA, GW170817_kilonova, Kasen2017, Cowperthwaite2017, Villar2017, Margalit2017}, and constraints from late-time observations of short gamma-ray bursts~\citep{Schroeder2020} suggest a maximum \ac{NS} mass of $\lesssim 2.1$--$2.7\,\Msun$. 

The upper end of the \acs{LMG} is motivated by mass determinations in \ac{XRB} systems. 
Though the lowest-mass \ac{BH} candidate to date is between $2.6$--$6.1\,\Msun$  ($95\%$ confidence; \citealt{Thompson2019}, see also \citealt{vandenHeuvel2020}), most \acp{BH} observed in the Galaxy have masses $\gtrsim 5\,\Msun$~\citep[see, e.g.,][]{Miller2015}. 
Selection effects may be affecting the observational sample~\citep{Kreidberg2012}, but it has been argued that such biases do not affect this broad picture~\citep[e.g.,][]{Ozel2010}. 

A gap (or lack thereof) in the mass spectrum of compact objects offers insights into the underlying \ac{SN} mechanism responsible for their formation. 
In particular, if instability growth and launch of the \ac{SN} proceed on rapid timescales ($\sim 10~\mathrm{ms}$ and $\sim 100~\mathrm{ms}$, respectively), stellar modeling and hydrodynamic simulations predict a dearth in remnant masses between $\sim 2$--$5\,\Msun$~\citep[see][]{Belczynski2012a,Fryer2012,Muller2016}. 
Alternatively, if instabilities are delayed and develop over longer timescales ($\gtrsim 200~\mathrm{ms}$), accretion can occur on the proto-\ac{NS} before the neutrino-driven explosion, and this gap would be filled~\citep{Fryer2012}. 
The mass distributions of \acp{NS} and \acp{BH} observed in our Galaxy provided initial evidence that this phenomenon proceeds on rapid timescales, and such prescriptions were therefore inherited by many rapid population studies for \ac{NS} and \ac{BH} systems~\citep[e.g.,][]{Dominik2012,Belczynski2014,Breivik2016a,Rodriguez2016a,Belczynski2016b,Mapelli2018a,Giacobbo2018b,Kremer2019b,Neijssel2019,Spera2019,Zevin2019b,DiCarlo2020,Rastello2020,Banerjee2020}. 
However, recently discovered Galactic compact objects with mass estimates that extend inside the \acs{LMG}~\citep[e.g.,][]{Thompson2019,Wyrzykowski2020} and population fits to known Galactic \acp{BNS}~\citep{Vigna-Gomez2018} raise tension with this interpretation; additional observations could further constrain \ac{SN} physics~\citep[e.g.,][]{Breivik2019}. 

GW190814 offers an unprecedented probe into the gap of compact object masses between \acp{BH} and \acp{NS}. 
The mass of the binary's secondary component is ${m_2 = \SecondaryMassNinety}$ ($90\%$ credibility), making it the heaviest \ac{NS} or lightest \ac{BH} ever identified in a compact-object binary. 
The secondary of GW190814 has the potential to provide insights into the \ac{SN} explosion mechanism since it is a relatively clean probe of the compact object's mass at birth; even if the lighter component of GW190814 was the first-born compact object, only small amounts of accretion are possible over the evolutionary timescale of its more massive companion.
Additionally, the more massive primary component is ${m_1\,\PrimaryMassApprox}$, making this the most asymmetric compact binary discovered, with a mass ratio of ${q = m_2/m_1 \MassRatioApprox}$~\citep[see][]{GW190412}. 
Such low mass ratios are predicted to be rare in both the canonical isolated binary evolution~\citep[e.g.,][]{Dominik2012, Giacobbo2018b, Kruckow2018, Spera2019} and dynamical formation~\citep[e.g.,][]{Clausen2013, Rodriguez2019, ArcaSedda2020} scenarios for compact-binary formation. 

We investigate compact-object formation in the \acs{LMG} and the formation of highly asymmetric compact binaries. 
We focus on the formation of GW190814-like systems using various standard assumptions for binary evolution, and whether population models can produce such systems while being consistent with the empirical merger rates of different compact-binary populations. 
We do not make specific alterations to our population models to preferentially form GW190814-like systems, but instead explore how uncertain model parameters affect the formation rate and properties of such systems. 
In Section~\ref{sec:populations}, we give an overview of our population models and present an updated prescription for determining remnant masses. 
We then discuss our model results in Section~\ref{sec:results}, including the merger rates and formation pathways of GW190814-like systems. 
We explain implications for binary evolution physics in Section~\ref{sec:conclusions}. 
Throughout we assume solar metallicity of ${Z_\odot=0.017}$~\citep{Grevesse1998} and \textit{Planck 2015} cosmological parameters of ${H_0 = 68~\mathrm{km\,s}^{-1}\,\mathrm{Mpc}^{-1}}$, ${\Omega_\mathrm{m} = 0.31}$ and ${\Omega_{\Lambda} = 0.69}$~\citep{PlanckCollaboration2016}.

\begin{figure*}
    \centering
    \includegraphics[width=\textwidth]{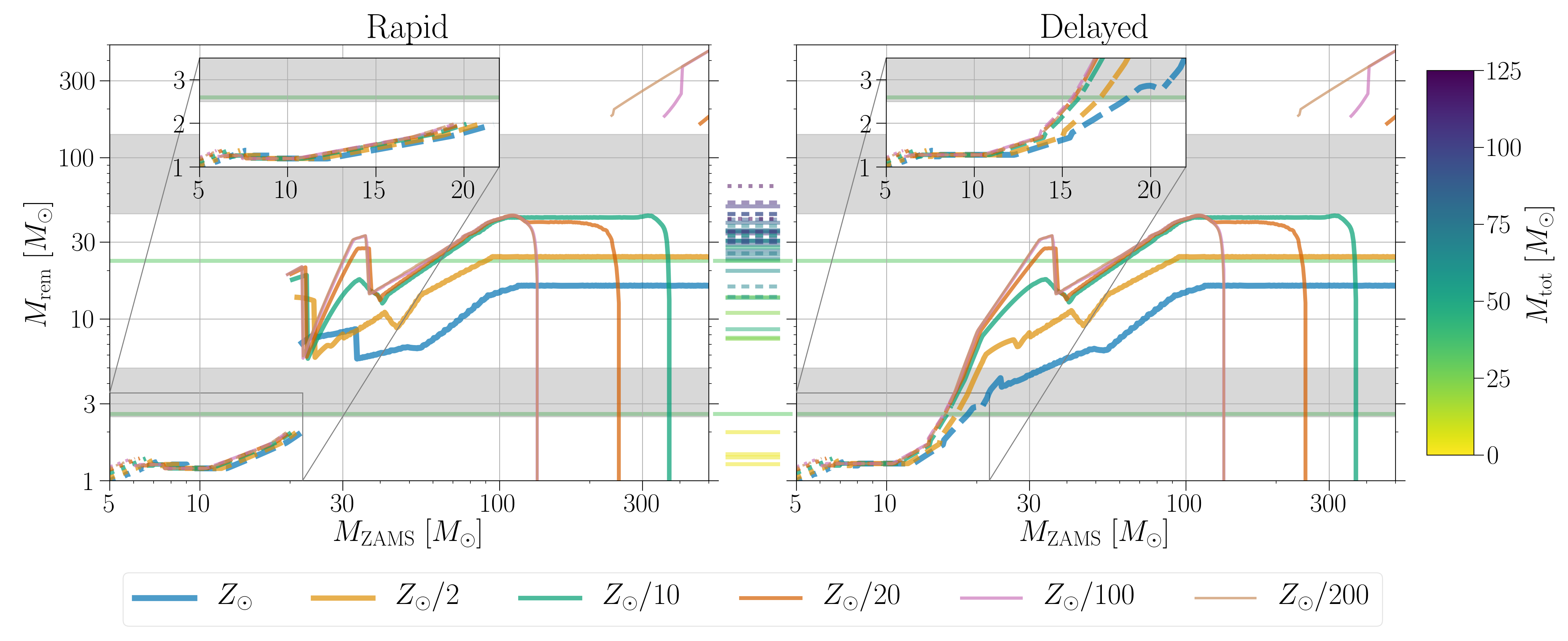}
    \caption{Relation between ZAMS mass and remnant mass for single stars in our models across a range of metallicities, using the Rapid (\emph{left}) and Delayed (\emph{right}) \ac{SN} mechanisms for determining compact-object masses. 
    White dwarfs, \acp{NS}, and \acp{BH} are distinguished by dotted, dashed, and solid lines, respectively; at low metallicities, the mass spectrum continues above the upper mass gap for high ZAMS masses. 
    The approximate lower and upper mass gaps are highlighted with gray bands. 
    Horizontal lines between the panels show the median component masses of compact binaries identified via \acp{GW} in the first catalog where they were reported with an astrophysical probability of ${p_\mathrm{astro}\,>\,0.5}$: the LIGO Scientific Collaboration and Virgo Collaboration catalog~\citep[solid lines;][]{GWTC1,GW190425,GW190412,GW190814}, the Institute for Advanced Studies catalog~\citep[dashed lines;][]{Venumadhav2019,Venumadhav2020} and the Open Gravitational-wave Catalog~\citep[dotted lines;][]{Nitz2019a,Nitz2020}; the components of GW190814 are also shown in the left and right panels and inset panels as a light green horizontal line. 
    Both components in an individual system are colored by their total mass. }
    \label{fig:Mzams_Mrem}
\end{figure*}

\section{Population Models}\label{sec:populations}

We use the rapid binary population synthesis code \texttt{COSMIC}~\citep{Breivik2020} to examine the properties and rates of compact binaries.\footnote{\href{https://cosmic-popsynth.github.io/}{cosmic-popsynth.github.io} (Version 3.3)} 
We investigate the impact of initial binary properties, efficiency of \ac{CE} evolution, survival in the \ac{CE} phase, the determination of remnant masses, and \ac{SN} natal kick prescriptions. 
We describe \texttt{COSMIC} and our model assumptions in Appendix~\ref{app:pop_models}, and our model variations are summarized in Table~\ref{tab:table}. 
Below we highlight updates to \texttt{COSMIC} that are pertinent to this work.

\subsection{Remnant Mass Prescription}\label{subsec:remnant_mass}

We follow the remnant mass prescriptions from \citet{Fryer2012} for determining the baryonic mass of the proto-compact object. 
The two prescriptions from \citet{Fryer2012}, \emph{Rapid} and \emph{Delayed}, are used to map the results of hydrodynamical simulations to rapid population synthesis. 
The two models differ in their assumed instability growth timescale (${\sim 10~\mathrm{ms}}$ and ${\sim 200~\mathrm{ms}}$ for Rapid and Delayed, respectively), with the Rapid model naturally leading to a low-mass gap. 

We make a small, but important, change when determining the final gravitational mass from the baryonic mass of the proto-compact object; further motivation and details are in Appendix~\ref{app:remnant_mass}. 
Rather than assuming a fixed fractional mass loss of the \emph{total} pre-\ac{SN} mass for \acp{BH} as in \citet{Fryer2012}, we cap the mass loss due to  neutronization to $10\%$ of the maximum mass assumed for the iron core, since hydrodynamical simulations show that the mass loss from neutrino emission is $\sim 10\%$ of the iron-core mass rather than the total baryonic mass of the compact-object progenitor~(C.\ Fryer 2020, private communication). 
Combining this criterion with the baryonic-to-gravitational mass prescription from \citet{Lattimer1989} gives
\begin{equation}
    M_\mathrm{grav} = 
    \begin{cases}
    \displaystyle \frac{20}{3} \left[(1 + 0.3 M_\mathrm{bar})^{1/2} - 1\right]& \Delta M\leq 0.1\,m_\mathrm{Fe}^\mathrm{max}\\
    M_\mathrm{bar}-0.1\,m_\mathrm{Fe}^\mathrm{max}& \mathrm{otherwise}\\
    \end{cases},
\end{equation}
where $\Delta M = M_\mathrm{bar}-M_\mathrm{grav}$ and $m_\mathrm{Fe}^\mathrm{max}$ is the maximum possible mass assumed for the iron core, which we set to $5\,\Msun$. 
This upper limit on the \ac{BH} mass loss when converting from baryonic to gravitational mass is slightly larger than, though qualitatively similar to, that in the procedure in \cite{Mandel2020a}. 
Whether the compact remnant is an \ac{NS} or a \ac{BH} can then be determined by comparing $M_\mathrm{grav}$ to $m_\mathrm{NS}^\mathrm{max}$. 
We show the relation between \ac{ZAMS} mass and remnant mass for both our \ac{SN} prescriptions in Figure~\ref{fig:Mzams_Mrem}.

\subsection{Merger Rates and Astrophysical Populations}\label{subsec:merger_rates}

Local merger rate densities can be determined from population synthesis to directly compare population predictions with the empirical rates measured by LIGO--Virgo. 
We calculate local merger rates of different compact binary populations in a similar manner to \cite{Giacobbo2018b} and \cite{Spera2019}, as described in detail in Appendix~\ref{app:merger_rates}.
For compact binary coalescence class $i$, the local merger rate (the merger rate at redshifts ${z \leq z_\mathrm{loc}} = 0.01$) across all metallicity models $j$ is
\begin{equation}
    \mathcal{R}_{\mathrm{loc},\,i} = \frac{1}{t_\mathrm{l}(z_\mathrm{loc})} \int_{0}^{z_\mathrm{max}} \psi(z) \sum_{j}p(Z_j | z) f_{\mathrm{loc},\,i}(z,Z_j) \frac{\mathrm{d} t_\mathrm{l}}{\mathrm{d}z} \mathrm{d}z, 
\end{equation}
where $t_\mathrm{l}$ is the lookback time, $\psi(z)$ is the star formation rate density, $p(Z_j | z)$ is the likelihood of metallicity $Z_j$ at redshift $z$, $f_{\mathrm{loc},\,i}(z,Z_j)$ is the mass fraction of systems born at redshift $z$ with metallicity $Z_j$ that merge in the local universe, and $z_\mathrm{max} = 15$ is the maximum redshift we consider for binary formation. 

To get a representative astrophysical population of compact binary mergers, we combine the information from across metallicities for each population model. 
We assume that all systems merge at the median measured redshift of GW190814 (${z\,\EventRedshift}$).\footnote{The low merger redshift of GW190814 makes these results a good proxy for local mergers in general.}  
The delay time of each system thus provides its formation time and formation redshift. 
We then give a weight to each system based on the mass-weighted star formation weight (Equation~\ref{eq:sfr}) at the formation redshift, the metallicity distribution (Equation~\ref{eq:met}) at the formation redshift, the relative formation efficiency for the double compact object population in question in the particular metallicity model, and the relative number of systems formed in a particular metallicity model. 
Drawing a subset of systems based on these weights gives a representative sample at the merger redshift of GW190814.

\section{Results}\label{sec:results}

\begin{figure}
    \centering
    \includegraphics[width=0.46\textwidth]{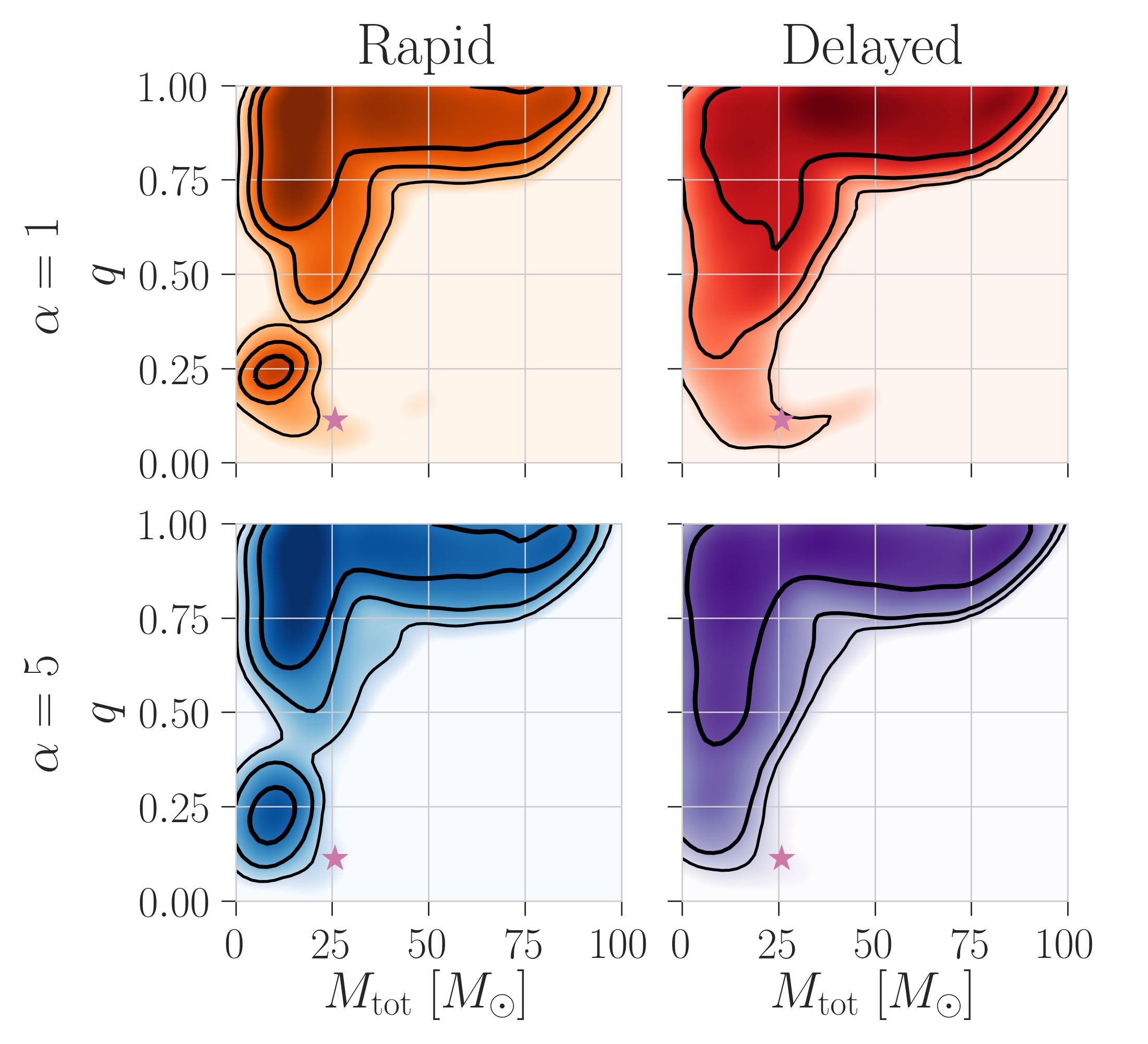}
    \caption{Distributions of total mass $M_\mathrm{tot} = m_1+m_2$ and mass ratio ${q =m_2/m_1}$ (${m_2 \leq m_1}$) of astrophysical populations for models varying the \ac{CE} efficiency $\alpha$ and remnant-mass prescription.
    For the other model variations, we use initial conditions from \citet{Sana2012}, the bimodal natal-kick model, and the Pessimistic assumption for \ac{CE} survival (see Appendix~\ref{app:model_assumptions}). 
    We include all systems with at least one component having ${m\,>\,3\,\Msun}$. 
    Systems are drawn from metallicity models based on the star formation history and metallicity evolution outlined in Section~\ref{subsec:merger_rates}, assuming they merge at the redshift of GW190814. 
    Densities are log-scaled, with contours containing $90\%$, $99\%$, and $99.9\%$ of systems. 
    GW190814's $M_\mathrm{tot}$ and $q$ are shown with a pink star; error bars are present, but are smaller than the marker. 
    }
    \label{fig:Mtot_q_credible}
\end{figure}

We explore the properties and rates of compact binaries in our population models, focusing on how standard population modeling and variations of physical assumptions inherent to binary stellar evolution impact the formation rate of asymmetric-mass binaries and mergers with components residing in the \acs{LMG}, especially GW190814-like systems. 
Given the uncertainty in $m_\mathrm{NS}^\mathrm{max}$, we show combined distributions for all systems with at least one component having a mass ${> 3\,\Msun}$ unless otherwise specified.

\subsection{Probing the Regime of Low Mass Ratio}

We find that merger rates drop precipitously as component masses become more asymmetric, in agreement with many other population synthesis studies~(e.g., \citealt{Dominik2012,Kruckow2018,Mapelli2018a,Neijssel2019}, though see also \citealt{Eldridge2016,Eldridge2017}). 
Figure~\ref{fig:Mtot_q_credible} shows the distribution of mass ratio and total mass in a subset of our populations, with contour lines marking $90\%$, $99\%$, and $99.9\%$ of the population. 
Mass ratios are concentrated near unity for large total masses; since the pair instability process limits the maximum mass of \acp{BH}~\citep[e.g.,][]{Woosley2017,Farmer2019,Marchant2019}, the degree of possible asymmetry decreases as a function of total mass. 
For systems with lower total mass, the mass ratio distribution extends to more asymmetric configurations, reaching down to ${q \lesssim 0.1}$. 

We do not see a strong difference in the distributions of total mass and mass ratio when varying the efficiency of \ac{CE} ejection. 
However, for the Rapid \ac{SN} mechanism we find fewer mass ratios near ${q \sim 0.4}$, and an island at lower mass ratios, whereas there is a continuum for the Delayed \ac{SN} mechanism. 
This is a byproduct of the \acs{LMG} that is inherent to the Rapid prescription; since this gap extends from ${\sim 3}$--${6\,\Msun}$, a system with ${M_\mathrm{tot}=18\,\Msun}$ cannot have a mass ratio between ${\sim 0.2}$--${0.5}$. 
However, even with the Delayed \ac{SN} mechanism we find ${\sim 90\%}$ of systems to have mass ratios of ${q\,\DelayedqTenPercentApprox}$. 
Regardless of our model assumptions, we find GW190814's mass ratio and total mass to be an outlier, lying close to the \EventOutlierStatus contour for our populations.

\begin{figure}[t!]
    \centering
    \includegraphics[width=0.46\textwidth]{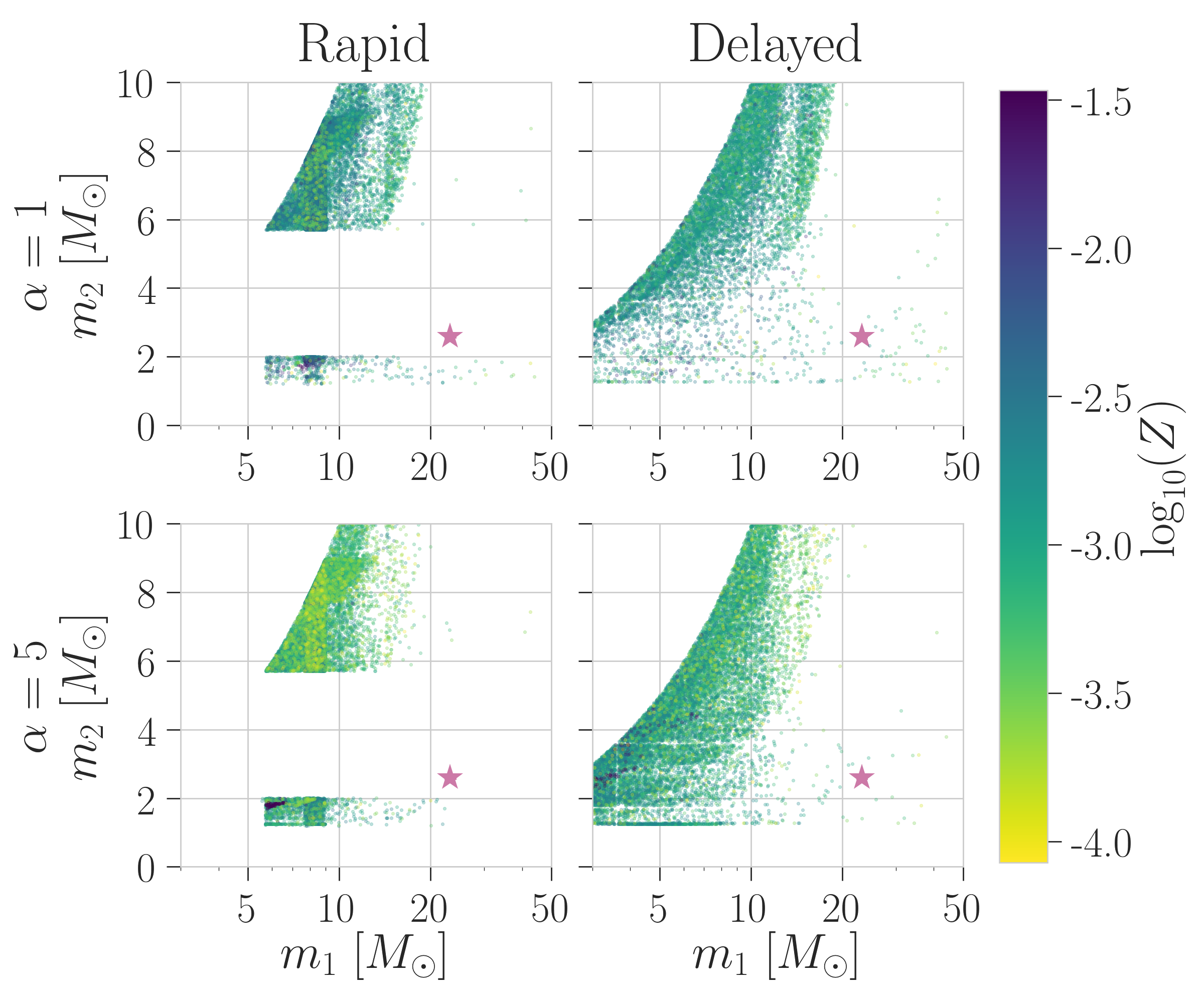}
    \caption{Primary mass $m_1$ and secondary mass $m_2$ for the same models as in Figure~\ref{fig:Mtot_q_credible}. 
    Systems are colored by their metallicity. 
    GW190814's component masses are shown with a pink star; error bars are again smaller than the marker. 
    }
    \label{fig:m1_m2_scatter}
\end{figure}

\subsection{Populating the Lower Mass Gap}

The impact of the \acs{LMG} is more apparent when examining systems' component masses. 
Figure~\ref{fig:m1_m2_scatter} shows the primary and secondary masses of systems merging at the redshift of GW190814 for a subset of our populations. 
Systems are more sparse when moving away from equal mass.  
Although rare, we do find systems matching GW190814's component masses when using the Delayed prescription, which naturally populates the \acs{LMG}. 
However, it is \emph{impossible} to form GW190814-like systems in our models using the Rapid \ac{SN} prescription. 

\begin{figure}[t!]
    \centering
    \includegraphics[width=0.46\textwidth]{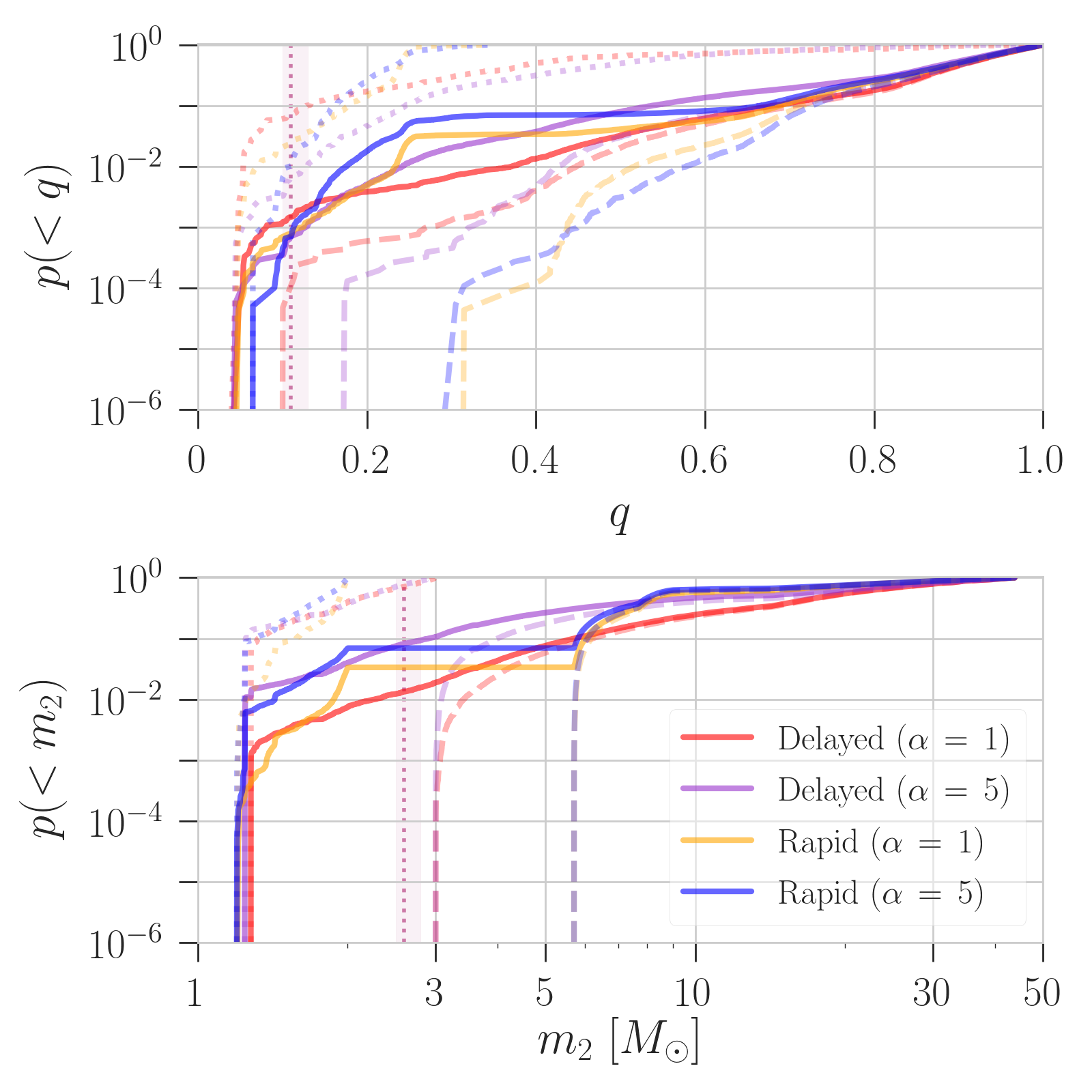}
    \caption{Cumulative distribution function for the mass ratio $q$ and secondary mass $m_2$ for the same models presented in Figure~\ref{fig:Mtot_q_credible}. 
    Solid lines show the combined population for all systems containing at least one component with ${m\,>\,3\,\Msun}$, whereas dashed and dotted lines show the corresponding cumulative distribution functions for the \emph{BBH} (${m_1 > 3\,\Msun,\ m_2 > 3\,\Msun}$) and \emph{NSBH} (${m_1 > 3\,\Msun,\ m_2 \leq 3\,\Msun}$) populations, respectively. 
    The dotted pink line and shaded region show the median and $90\%$ credible interval for GW190814. 
    }
    \label{fig:q_m2_cdf}
\end{figure}

In Figure~\ref{fig:q_m2_cdf}, we show cumulative distributions for the mass ratio and secondary mass in our populations. 
In the full population, we find that one in \FullPopMassRatiosLikeEvent systems have a mass ratio similar to GW190814 or lower (${q\,\leq\,\MassRatioHigh}$). For systems with a secondary mass ${\leq\,3\,\Msun}$, \NSBHMassRatiosLikeEvent of systems have a mass ratio of ${q\,\leq\,\MassRatioHigh}$, though this drops to \NSBHMassRatiosLikeEventEfficientCE when a more efficient \ac{CE} is assumed, which is qualitatively similar to findings from other population synthesis work \citep[e.g.,][]{Giacobbo2018b}. 
For systems with a secondary mass ${>\,3\,\Msun}$, the mass ratio distribution deviates significantly when assuming the Rapid \ac{SN} mechanism compared to the Delayed mechanism. 
We find \BBHMassRatiosLikeEvent of these systems have mass ratios of ${q\,\leq\,0.12}$ with a Delayed \ac{SN} mechanism, whereas these systems are nonexistent when assuming a Rapid \ac{SN} mechanism.
The \acs{LMG} can be seen in the bottom panel of Figure~\ref{fig:q_m2_cdf} as a plateau in the Rapid models as a function of $m_2$; the Delayed models, which populate the gap, have a more gradual buildup. 
For Delayed models, GW190814's secondary mass lies at about the \FullPopMtwoLikeEventAlphaFive percentile of the full population for ${\alpha=5}$, and drops to about the \FullPopMtwoLikeEventAlphaOne percentile for ${\alpha=1}$.

\subsection{Compact Binary Merger Rates}\label{subsec:rate_predictions}

\begin{figure}[t!]
    \centering
    \includegraphics[width=0.46\textwidth]{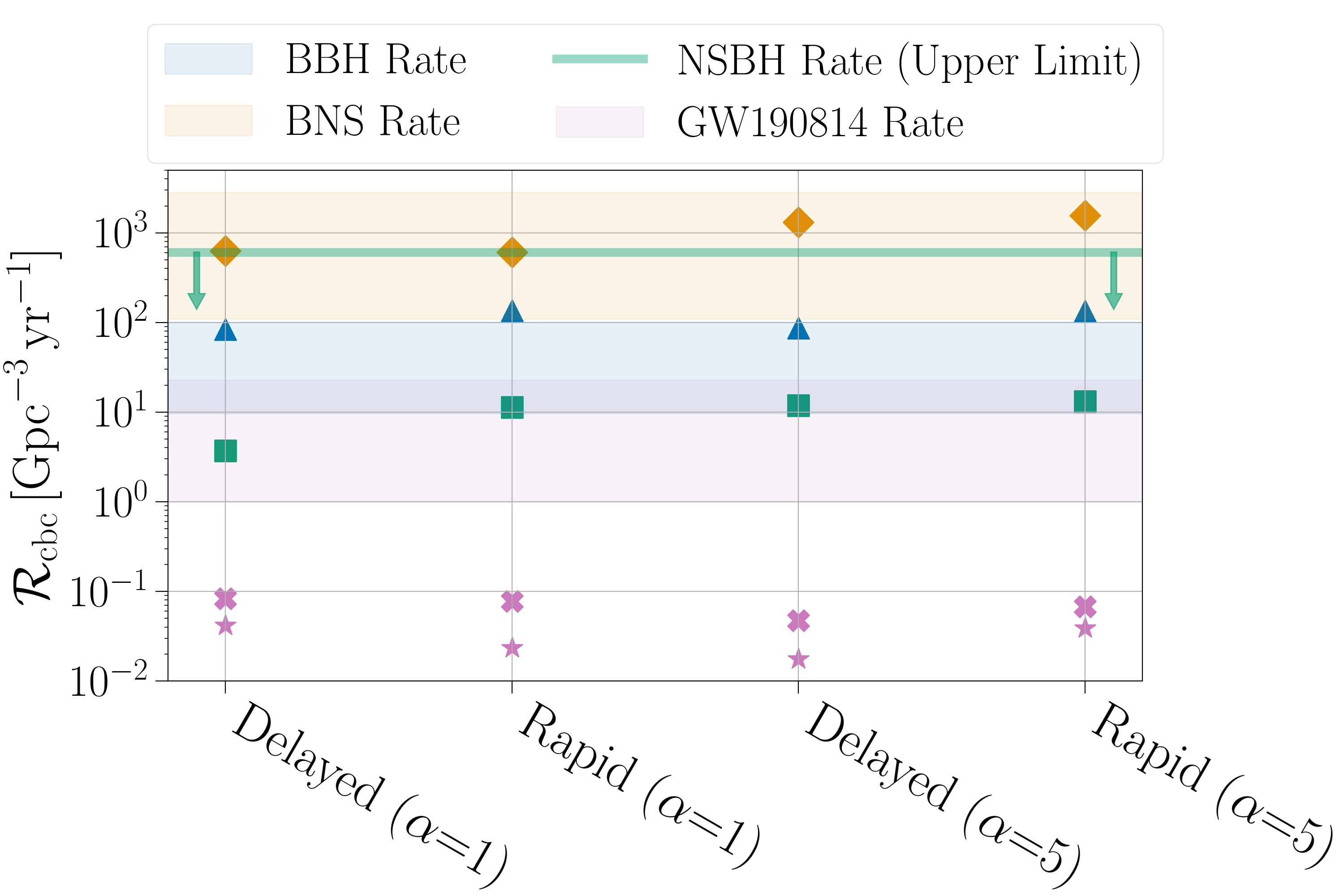}
    \caption{Local merger rates for models where we vary the \ac{CE} efficiency and remnant mass prescription. 
    The blue and orange shaded regions mark the $90\%$ credible level for the current empirical \ac{BBH}~\citep{GWTC1} and \ac{BNS}~\citep{GW190425} merger rates from the \ac{LVC}, respectively, the green line marks the $90\%$ credible upper limit for the \ac{LVC} \ac{NSBH} rate~\citep{GWTC1}, and the pink shaded region marks the $90\%$ credible single-event rate for GW190814-like systems~\citep{GW190814}. 
    Matching colored symbols mark the \ac{BBH} (blue triangles), \ac{BNS} (orange diamonds), and \ac{NSBH} (green squares) merger rates from our models, assuming ${m_\mathrm{NS}^\mathrm{max} = 3\,\Msun}$. 
    Pink symbols mark the merger rates from our models for systems with ${q \leq 0.2}$ and ${M_\mathrm{tot}/\Msun \geq 20}$ (crosses), and ${0.06\leq q \leq 0.16}$ and ${20 \leq M_\mathrm{tot}/\Msun \leq 30}$ (stars). 
    }
    \label{fig:model_rates}
\end{figure}

Merger rates are a useful diagnostic for comparing predictions of population synthesis modeling to the empirical merger rate estimated by the \ac{LVC}. 
Figure~\ref{fig:model_rates} shows local merger rates of different compact binary populations for four variations of model assumptions. 
To compare with \ac{LVC} rates, we assume that compact objects with masses $\leq 3\,\Msun$ are \acp{NS}, and those with masses $> 3\,\Msun$ are \acp{BH}. 
For the four models examined, we find \ac{BBH}, \ac{NSBH}, and \ac{BNS} merger rates to be consistent with the measured \ac{LVC} rate (bands for \acp{BBH} and \acp{BNS}, upper limit for \acp{NSBH}). 
We do not expect exact agreement in rates, since the \ac{LVC} results were calculated using mass distributions that are different from our populations.

The single-event rate for GW190814-like systems is $1$--$23$\,Gpc$^{-3}$\,yr$^{-1}$~\citep[90\% credible level;][]{GW190814}, and is shown with a pink band in Figure~\ref{fig:model_rates}. 
To compare our model predictions with the empirical rate, we choose two approximations for identifying GW190814-like systems in our models: a \emph{Narrow} GW190814-like rate where we choose systems with ${0.06 \leq q \leq 0.16}$ and ${20 \leq M_\mathrm{tot}/\Msun\,\leq\,30}$ (pink stars) and a \emph{Broad} GW190814-like rate where we choose systems with ${q \leq\,0.2}$ and ${M_\mathrm{tot}/\Msun \geq 20}$. 
In both cases, we find the local merger rate of GW190814-like systems to be over an order of magnitude lower than the empirical GW190814 rate. 
For example, in our model with a Delayed \ac{SN} mechanism and $\alpha=1$, we find a local merger rate of \DelayedAlphaOneBroadRate for our Broad GW190814-like assumption and \DelayedAlphaOneNarrowRate for our Narrow GW190814-like assumption; for a more efficient \ac{CE}, the Broad and Narrow rates drop by factors of \DelayedAlphaFiveBroadDrop and \DelayedAlphaFiveNarrowDrop, respectively. 
Merger rates for our other model variations are presented in Table~\ref{tab:table}.

\subsection{Formation of GW190814-like systems}\label{subsec:formation_pathways}

\begin{figure}[t!]
    \centering
    \includegraphics[width=0.46\textwidth]{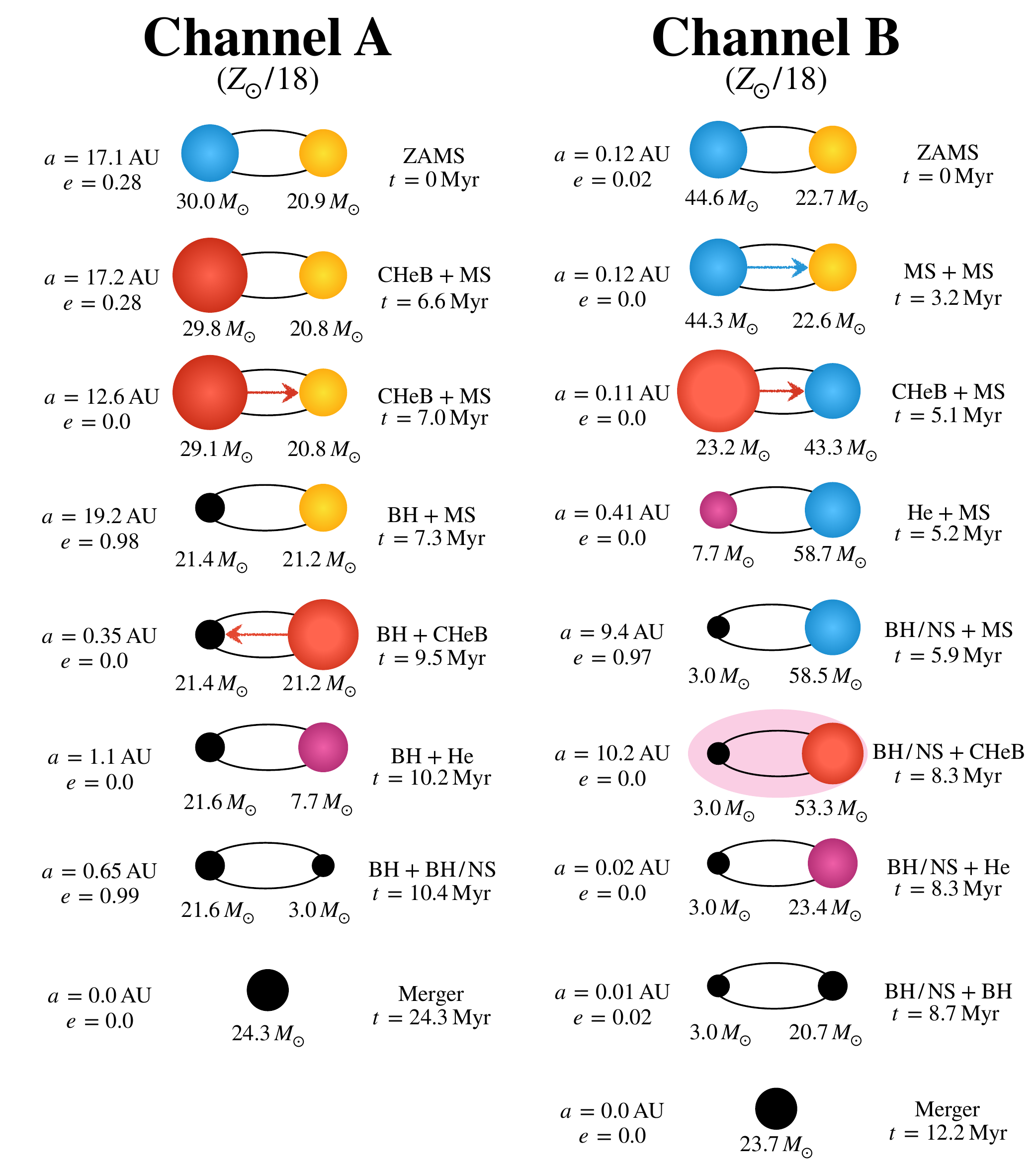}
    \caption{Evolutionary diagrams of two systems following the dominant pathways for forming GW190814-like systems in our models, as described in Section~\ref{subsec:formation_pathways}. 
    In Channel A, the more massive star at \ac{ZAMS} becomes the larger compact object, whereas in Channel B the system undergoes a mass inversion and the more massive star at \ac{ZAMS} evolves into the lighter compact object. 
    Stable mass transfer is denoted by an arrow from the donor to the accretor, and a \ac{CE} phase in Channel B by the pink oval surrounding the two stars. 
    }
    \label{fig:formation_channels}
\end{figure}

We identify two main channels for forming GW190814-like systems through isolated binary evolution in our population models. 
These can be broadly categorized as Channel A, where the primary (more massive) star at \ac{ZAMS} becomes the more massive \ac{BH} component in the compact binary, and Channel B, where the primary star at \ac{ZAMS} becomes the less massive object in the compact binary (either an \ac{NS} or a \ac{BH}). 
Figure~\ref{fig:formation_channels} shows evolutionary diagrams for examples from these general channels. 

In Channel A, the binary typically starts as a \ChannelAPrimary primary and \ChannelASecondary secondary at \ac{ZAMS}. 
The binary evolves without interaction during the main sequence of the primary. 
At core helium burning, the primary overflows its Roche lobe. 
Depending on the mass ratio at the time, the mass transfer will proceed either stably or unstably, in the latter case triggering a \ac{CE} phase~\citep{Taam2000}. 
The primary then directly collapses into a $\sim 20\,\Msun$ \ac{BH}. 
As the secondary crosses the Hertzsprung gap, it overflows its Roche lobe and proceeds through highly non-conservative mass transfer. 
Since the mass ratio between the donor star and the already-formed \ac{BH} is close to unity, this phase of mass transfer typically proceeds stably. 
The naked helium-star does not proceed through another phase of mass transfer and becomes an \ac{NS} following its \ac{SN}. 
These systems generally have large orbital separations at double compact-object formation, and thus to merge within a Hubble time the newly formed \ac{NS} needs to be kicked into a highly eccentric orbit, typically with a post-\ac{SN} eccentricity of \ChannelAEccentricity. 

In Channel B, the binary starts with a \ac{ZAMS} mass ratio of \ChannelBMassRatio. 
The primary fills its Roche lobe while on the main sequence. 
This phase of stable mass transfer donates a significant amount of material to the secondary, leading to a mass inversion where the lighter star at \ac{ZAMS} becomes the more massive star. 
The initially more massive star forms the lighter compact object before the secondary leaves the main sequence. 
Due to the large mass asymmetry between the already-formed \ac{NS} and the now more massive secondary, when the secondary evolves into a giant and overflows its Roche lobe, mass transfer proceeds unstably and initiates a \ac{CE}. 
The binary significantly hardens during the \ac{CE} phase, and the second star directly collapses into a \ac{BH}. 
This channel for forming GW190814-like systems was also identified in \cite{Mandel2020b}, though in our case it does not necessitate an Optimistic \ac{CE} scenario, and we still find systems from this channel when binaries with unstable mass transfer from Hertzsprung gap donors are assumed to merge. 

At low metallicities, these channels operate with similar probability. 
For metallicities of ${Z\,\leq\,Z_\odot/30}$, we find that \ChannelAFracLowMet of systems in our simulated population with ${0.06\,\leq\,q\,\leq 0.16}$ and ${20\,\leq\,M_\mathrm{tot}/\Msun\,\leq\,30}$ lead to the primary \ac{ZAMS} star becoming the heavier compact object, whereas \ChannelBFracLowMet lead to the secondary \ac{ZAMS} star becoming the heavier compact object. 
At higher metallicities (${Z\,>\,Z_\odot/30}$) Channel A becomes dominant, with \ChannelAFracHighMet of systems proceeding through Channel A and \ChannelBFracHighMet through Channel B. 
As the metallicity increases to ${\gtrsim\,Z_\odot/8}$, GW190814-like systems no longer form since \ac{BH} masses are suppressed due to line-driven winds. 
Since there are only a small number of GW190814-like systems in our models, the channels presented here and their relative likelihood only offer a broad picture of how systems with properties similar to GW190814 form. 

\pagebreak
\section{Discussion and Conclusions}\label{sec:conclusions}

GW190814 is a \ac{GW} event that challenges compact binary population modeling and places new constraints on the physics of massive-star evolution. 
We explore the formation of compact binary mergers with highly asymmetric masses and components residing in the \acs{LMG}. 
We find that systems with properties similar to GW190814 \emph{can} form through isolated binary evolution; however, the predicted formation rates of such systems are \emph{an order of magnitude lower} than the empirical single-event rate when considering models that match the observed rates for other compact binary populations. 
The mass of GW190814's secondary lies in the dearth of compact objects between the heaviest \acp{NS} and lightest \acp{BH}, and \emph{if} it is the result of isolated binary evolution, requires that instability growth and \ac{SN} launch proceed on longer timescales than typically assumed. 

The most massive Galactic \acp{NS} have low-mass stellar companions, allowing for stable mass accretion on significant timescales. 
Thus, the masses of these \acp{NS} are not a direct probe of their masses at formation. 
GW190814's lower-mass component likely had minimal accretion after formation: its mass at merger is indicative of its birth mass. 
Even if there was a mass inversion in the system and the secondary component of GW190814 was the first-born compact object (Channel B in Section~\ref{subsec:formation_pathways}), the amount of material it could feasibly accrete is limited by the evolutionary timescale of its massive companion (which goes on to form a $\gtrsim 20\,\Msun$ \ac{BH}). 
The amount accreted by an \ac{NS} or \ac{BH} at the Eddington limit is
\begin{equation}
    \Delta M_\mathrm{Edd} \approx 0.03 \left(\frac{R}{10~\mathrm{km}}\right) \left(\frac{t}{1~\mathrm{Myr}}\right) \Msun,
\end{equation}
where $R$ is the radius of the compact object~\citep[e.g.,][]{Cameron1967}. 
Thus, even for accretion timescales of ${\mathcal{O}(1\,\mathrm{Myr})}$ and reasonable \ac{NS} radii~\citep{GW170817_EoS,Miller2019,Riley2019}, the amount of mass that the lighter compact object could accrete is ${\approx\,0.03\,\Msun}$, far too low to bridge the gap between the canonical ${1.3\,\Msun}$ \ac{NS} mass and the ${\SecondaryMassApprox}$ secondary of GW190814. 
While there is evidence of super-Eddington accretion in ultraluminous X-ray sources~\citep[e.g.,][]{Bachetti2014}, it is unlikely that this could increase the mass of the GW190814's lower-mass component significantly; the post-hydrogen exhaustion lifetime for the progenitor star of a ${\PrimaryMassApprox}$ \ac{BH} is only ${\sim 0.5~\mathrm{Myr}}$, and mass transfer between the high-mass companion star and the already-formed compact object would almost assuredly be unstable due to the system's large mass ratio. 
Thus, even at $10$ times the Eddington rate, GW190814's lower-mass component would not accrete more than ${\mathcal{O}(0.1\,\Msun)}$ during the remaining lifetime of its stellar companion. 

As a clean probe of the natal mass of a compact object in the \acs{LMG}, GW190814 is an exquisite system for constraining the \ac{SN} mechanisms that impact remnant masses. 
The secondary's mass is inconsistent with instability growth and \ac{SN} launch on rapid timescales~\citep[${t\,\sim\,10~\mathrm{ms}}$; e.g.,][]{Fryer2012,Muller2016}; if GW190814's source formed from isolated binary evolution, it favors \ac{SN} launch models where instabilities develop over a longer timescale (${t \sim\,100~\mathrm{ms}}$). 
As we only consider two \ac{SN} models, we cannot place a lower limit on the instability growth timescale. 
More detailed hydrodynamical simulations investigating instability growth, \ac{SN} launch, and how they connect to compact remnant masses~\citep[e.g.,][]{Ertl2020,Patton2020} will be needed to determine a lower limit on the timescales necessary to produce systems with component masses in the \acs{LMG}, and whether there is a critical growth timescale that leads to a populated \acs{LMG}. 
Semi-analytic prescriptions and fitting formulae based on these detailed models can then be adopted by rapid population synthesis, using either deterministic or probabilistic~\citep[e.g.,][]{Mandel2020a} approaches. 

The combination of a highly asymmetric binary with a low-mass secondary is predicted to be rare in most rapid binary population studies~\citep[e.g.,][]{Dominik2012, Giacobbo2018b, Kruckow2018, Mapelli2018a, Mapelli2019, Neijssel2019, Spera2019, Olejak2020}, though population modeling that allows for unlimited super-Eddington accretion onto \acp{BH} finds that compact binaries with $q \sim 0.1$ have rates comparable to near-equal-mass mergers~\citep{Eldridge2016,Eldridge2017} and may be able to better match the mass asymmetry of GW190814. 
Even when we consider \ac{SN} mechanisms that fill the \acs{LMG}, our predicted rate for GW190814-like systems is in tension with the empirical \ac{LVC} rate. 
There are many theoretical uncertainties in binary stellar evolution with complex correlations that strongly affect the rates and population properties of compact binary mergers~\citep[e.g.,][]{Barrett2018}, and we only choose a few to investigate. 
It is possible that variations in other uncertain physical prescriptions, such as the mass-transfer accretion rates, mass-transfer efficiency, the criteria for the onset of unstable mass transfer, and how each of these depends on the evolutionary stages of the stars involved, may help to alleviate the discrepancy between the merger rates of GW190814-like systems and those of other compact binary populations. 
Observations of compact binaries with unusual properties (such as GW190814) will be paramount in constraining uncertainties in this high-dimensional parameter space. 

This work focuses on the formation of systems with high mass ratios and component masses in the \acs{LMG} through canonical isolated binary evolution. 
Many other channels have been proposed for producing the compact binary mergers observed by LIGO--Virgo. 
Dynamical formation in dense stellar clusters such as globular clusters preferentially produces compact binaries with similar masses~\citep[e.g.,][]{Sigurdsson1993a}, and thus the formation of \acp{NSBH} and other compact binaries with highly asymmetric masses is predicted to be rare~\citep{Clausen2013,ArcaSedda2020,Ye2020}. 
While hierarchical mergers of \acp{NS} have been proposed as a means of populating the \acs{LMG}~\citep{Gupta2020}, this scenario is unlikely since heavier \acp{BH} dominate the dynamical interactions in clusters~\citep[e.g.,][]{Samsing2020,Ye2020}. 
The formation of compact binaries with highly asymmetric masses may be more prevalent in young star clusters \citep[e.g.,][]{DiCarlo2019,Rastello2020,Santoliquido2020}, but \cite{Fragione2020a} finds the merger rate of \ac{NSBH} systems in young massive and open clusters to be three orders of magnitude lower, similar to the predictions from old globular clusters. 
Other formation mechanisms have been explored for forming highly asymmetric compact binary mergers and mergers with components in the \acs{LMG}, such as hierarchical systems in the galactic field~\citep[e.g.,][]{Antonini2017a,Silsbee2017,Fragione2019,Safarzadeh2019c,Fragione2020b}, hierarchical systems in galactic nuclei with a supermassive \ac{BH} as the outer perturber~\citep[e.g.,][]{Antonini2012,Petrovich2017,Hoang2018,Fragione2019b,Stephan2019}, and in disks around supermassive \acp{BH} in active galactic nuclei~\citep[e.g.,][]{McKernan2019,Yang2019b}. 
However, the rates and formation properties from these channels are uncertain. 
Nevertheless, a full picture of compact-binary mergers will require consideration of all these channels and investigation of how physical prescriptions (such as the connection between the underlying \ac{SN} mechanism and remnant mass) jointly affect population properties, rates, and branching ratios across these channels~\citep[e.g.,][]{Stevenson2017,Talbot2017,Vitale2017a,Zevin2017b,ArcaSedda2020}.
The identification of bona fide \ac{NSBH} systems and other compact-binary mergers with highly asymmetric masses will further constrain the relative contribution of various formation channels and the underlying physics of these formation pathways. 

\acknowledgements
The authors thank Chris Fryer, Pablo Marchant, and Ilya Mandel for useful discussions, and we thank the anonymous referee for helpful suggestions that improved this paper. 
M.Z. acknowledges support from CIERA and Northwestern University. 
M.S. acknowledges funding from the European Union's Horizon 2020 research and innovation program under the Marie-Sk\l{}odowska-Curie grant agreement No.\ 794393.
C.P.L.B. is supported by the CIERA Board of Visitors Professorship. 
V.K. is supported by a CIFAR G+EU Fellowship and Northwestern University. 
This work used computing resources at CIERA funded by NSF grant No.\ PHY-1726951, and resources and staff provided for the Quest high performance computing facility at Northwestern University, which is jointly supported by the Office of the Provost, the Office for Research, and Northwestern University Information Technology. 

\software{\texttt{COSMIC}~\citep{Breivik2020}, \texttt{iPython}~\citep{ipython}, \texttt{Matplotlib}~\citep{matplotlib}, \texttt{NumPy}~\citep{numpy,numpy2}, \texttt{Pandas}~\citep{pandas}, \texttt{SciPy}~\citep{scipy}.}

\clearpage
\appendix

\section{Population Models}\label{app:pop_models}

\texttt{COSMIC}~\citep{Breivik2020} is based on the single-star fitting formulae from \cite{Hurley2000} and binary evolution prescriptions from \cite{Hurley2002}. 
Among many updates, \texttt{COSMIC} includes state-of-the-art physical prescriptions for stellar winds in massive stars~\citep{Vink2001} and stripped stars~\citep{Vink2005,Yoon2005}, multiple treatments for the onset~\citep{Belczynski2008,Claeys2014} and evolution~\citep{Claeys2014} of unstable mass transfer, multiple prescriptions for \ac{SN} natal kicks~\citep{Hobbs2005,Bray2016,Giacobbo2020} with special treatment for electron-capture \acp{SN}~\citep{Podsiadlowski2004} and ultra-stripped \acp{SN}~\citep{Tauris2015}, as well as mass loss and orbital evolution from pulsation pair instabilities and pair instability \acp{SN}~\citep{Woosley2017,Woosley2019,Marchant2019}. 
Furthermore, \texttt{COSMIC} includes a number of variations for how initial conditions are sampled~\citep{Sana2012,Moe2017}, which can significantly affect the properties and rates of compact binary populations. 
Rather than simulating a predetermined number of systems, \texttt{COSMIC} runs populations specifically targeted at particular configurations of stellar types (such as \acp{BNS} or \acp{BBH} that merge within a Hubble time) until properties of the target population (such as their masses and orbital periods at formation) have converged~\citep{Breivik2020}, thereby adequately exploring the tails of population distributions. 

\subsection{Model Assumptions}\label{app:model_assumptions}

We investigate five uncertain aspects of binary evolution physics. 
\begin{enumerate}
    \item Initial conditions (primary mass, mass ratio, orbital period, and eccentricity) are sampled either independently using the best-fit values from \citet{Sana2012} with a binary fraction of $0.7$, or using the correlated multidimensional distributions from \citet{Moe2017}. 
    In the multidimensional sampling, the binary fraction is determined based on the probability that a system with a given primary mass is in a binary. 
    \item \ac{CE} efficiency, which determines how easily the envelope is unbound from the system during a \ac{CE} phase, is parameterized as in \citet{Webbink1984} and \citet{deKool1990}. 
    We vary the efficiency parameter $\alpha$, using either $\alpha=1$ or a higher value of $\alpha=5$~\citep{Fragos2019,Giacobbo2019a}, and use a variable prescription for the envelope binding energy factor $\lambda$~\citep{Claeys2014}. 
    A higher \ac{CE} efficiency will lead to wider post-\ac{CE} binaries. 
    \item \ac{CE} survival is chosen to be either an Optimistic or a Pessimistic scenario. 
    In the Optimistic case, stars that overfill their Roche lobes on the Hertzsprung gap and proceed through unstable mass transfer are assumed to survive the \ac{CE} phase, whereas in the Pessimistic case these systems are assumed to merge~\citep[see][]{Belczynski2008}. 
    The Pessimistic scenario leads to significantly fewer compact binary mergers, particularly for \acp{BBH}. 
    \item Remnant masses are determined using the Rapid and Delayed prescriptions from \citet{Fryer2012}. 
    The Rapid prescription yields a mass gap between \acp{NS} and \acp{BH}, whereas the Delayed prescription fills this gap (Figure~\ref{fig:Mzams_Mrem}). 
    These prescriptions are updated as described in Section~\ref{subsec:remnant_mass} and Section~\ref{app:remnant_mass}. 
    \item \ac{SN} natal kicks are determined in two ways. 
    In the bimodal prescription, iron core-collapse \acp{SN} kicks are drawn from a Maxwellian distribution with a dispersion of $\sigma=265~\mathrm{km\,s}^{-1}$~\citep{Hobbs2005}, whereas electron-capture \acp{SN} and ultra-stripped \acp{SN} are given weaker kicks drawn from a Maxwellian distribution with a dispersion of $\sigma=20~\mathrm{km\,s}^{-1}$ (e.g., \citealt{Podsiadlowski2004,VanDenHeuvel2007,Tauris2015,Beniamini2016}, see \citealt{Breivik2020} for more details). 
    The second kick prescription uses the scaling based on compact-object mass and mass loss in \citet{Giacobbo2020}. 
\end{enumerate}
These represent only a few of the binary evolution parameters that can possibly affect the parameter distribution and merger rates of compact-binary populations. 
Besides the parameter variations described above, we anticipate that mass transfer conservation and the stellar-type specific criteria for the onset of unstable mass transfer will have the largest impact. 
In this study, we assume that mass transfer is limited to the thermal timescale of the accretor for stars and limited to the Eddington rate for compact objects, and that angular momentum is lost from the system as if the excess material is a wind from the accretor~\citep{Hurley2002}. 
The onset of unstable mass transfer is determined as in \citet{Belczynski2008} using critical mass ratios for a given stellar type: $q_\mathrm{crit}=3.0$ for H-rich stars ($k_\star =  1$--$6$), $q_\mathrm{crit}=1.7$ for helium main-sequence stars ($k_\star = 7$), $q_\mathrm{crit}=3.5$ for evolved helium stars ($k_\star = 8,\,9$), and $q_\mathrm{crit}=0.628$ for compact objects ($k_\star \geq 10$). 
A full exploration of parameter space is reserved for future work. 

\subsection{Remnant Mass Prescription}\label{app:remnant_mass}

Here, we provide more details regarding the updated remnant mass prescription used in this study. 
To determine the mass of compact remnants, we follow the Rapid and Delayed prescriptions described in \citet{Fryer2012}. 
These allow for the results of hydrodynamical simulations exploring the timescale of instability growth and launch of the \ac{SN} to be used directly in rapid population synthesis. 
Mass fallback is also accounted for in the determination of the baryonic mass of the proto-compact object. 
As in \cite{Giacobbo2020}, we adjust the initial mass of the proto-compact object to be $1.1\,\Msun$ rather than $1.0\,\Msun$, as this better reproduces the typical masses of \acp{NS} in the Galaxy. 

Following the determination of the baryonic mass of the remnant, the gravitational mass is calculated to account for neutronization in the collapsing core. 
In \cite{Fryer2012}, the gravitational mass of the remnant is calculated differently for \acp{NS} and \acp{BH}. 
For \acp{NS}, the gravitational mass is calculated according \cite{Lattimer1989} based on the neutrino observations of SN~1987A: 
\begin{equation}
    M_\mathrm{grav} = \frac{20}{3} \left[(1 + 0.3 M_\mathrm{bar})^{1/2} - 1\right],
\end{equation}
where $M_\mathrm{bar}$ is the pre-collapse baryonic mass calculated as in \cite{Fryer2012}. 
For \acp{BH} the mass reduction is assumed to be a fixed percentage of the proto-compact object's baryonic mass: 
\begin{equation}
    M_\mathrm{grav} = 0.9 M_\mathrm{bar}. 
\end{equation}
This leads to an increasing amount of mass loss when converting from baryonic to gravitational mass as a function of increasing \ac{BH} mass. 
Since the true maximum \ac{NS} mass is unknown and likely sensitive to other aspects of the proto-compact object such as rotation, the delineation between these two prescriptions is typically determined by an adjustable parameter for the maximum \ac{NS} mass: $m_\mathrm{NS}^\mathrm{max}$. 

There are two issues with this simple prescription that affect the compact-object mass spectrum. 
First, the final mass of a \ac{BH} remnant is a function of the \emph{total} pre-collapse baryonic mass of the proto-compact object, though neutronization is instead occurring in the iron core of the proto-compact object. 
Even for massive \acp{BH} and hot radiation-supported cores, the iron core mass is $\lesssim 5\,\Msun$~(C.\ Fryer 2020, private communication). 
Hydrodynamical simulations show that the mass loss from neutrino emission is $\sim 10\%$ of this core mass rather than the total baryonic mass of the \ac{BH} progenitor~(C. Fryer 2020, private communication). 
Second, using separate prescriptions for determining \ac{NS} and \ac{BH} gravitational mass leads to an artificial gap in the mass spectrum; this artificial gap is different than the \acs{LMG} and is apparent even when using the Delayed remnant mass prescription. 
For example, assuming $m_\mathrm{NS}^\mathrm{max}=2.5\,\Msun$ and $10\%$ mass loss when converting from baryonic to gravitational mass in \acp{BH}, the most massive \ac{NS} that can be formed is $2.5\,\Msun$ whereas the least massive \ac{BH} that can be formed is $2.7\,\Msun$. 

With these in mind, we update how the final gravitational mass of a compact object is determined: 
\begin{equation}
    M_\mathrm{grav} = 
    \begin{cases}
    \displaystyle \frac{20}{3} \left[(1 + 0.3 M_\mathrm{bar})^{1/2} - 1\right]& \Delta M\leq 0.1\,m_\mathrm{Fe}^\mathrm{max}\\
    M_\mathrm{bar}-0.1\,m_\mathrm{Fe}^\mathrm{max}& \mathrm{otherwise}\\
    \end{cases},
\end{equation}
where $\Delta M = M_\mathrm{bar}-M_\mathrm{grav}$ and $m_\mathrm{Fe}^\mathrm{max}$ is the maximum possible mass of the iron core, which we set to $5\,\Msun$. 
For $m_\mathrm{Fe}^\mathrm{max}=5\,\Msun$, the switchover in this conditional occurs at $\simeq\,3.1\,\Msun$. 
As shown in Figure~\ref{fig:Mzams_Mrem}, this update eliminates any artificial gaps in the mass spectrum between \acp{NS} and \acp{BH} when using the Delayed \ac{SN} mechanism.

\section{Local Merger Rates}\label{app:merger_rates}

The mass fraction of binaries that are born at redshift $z$ and merge as compact binaries in the local universe is 
\begin{equation}
    f_{\mathrm{loc},\,i}(z) = \frac{M_i(z;z_\mathrm{merge}<z_\mathrm{loc})}{M_\mathrm{samp}},
\end{equation}
\noindent where $M_\mathrm{samp}$ is the total stellar mass sampled in the simulation, $i$ represents the class of compact binary merger (\ac{BNS}, \ac{NSBH}, \ac{BBH}, etc.), $M_i$ is the stellar mass that leads to merger type $i$, and $z_\mathrm{loc}$ is the maximum redshift that we consider for local mergers, which we set to $z_\mathrm{loc}=0.01$. 
\texttt{COSMIC} accounts for the total mass sampled in $M_\mathrm{samp}$, incorporating both the binary fraction and the mass contribution from lower-mass stars that do not lead to compact-binary formation. 
The number of mergers per unit volume that form in the redshift interval $[z,z+\Delta z]$ and merge in the local universe is thus
\begin{equation}
    \Delta \mathcal{N}_{\mathrm{loc},\,i}(z)= \psi(z) f_{\mathrm{loc},\,i}(z) \frac{\mathrm{d}t_\mathrm{l}}{\mathrm{d}z} \Delta z,
\end{equation}
where $\psi(z)$ is the star formation rate density and $t_\mathrm{l}(z)$ is the lookback time at redshift $z$.
We use \citet{Madau2017a} for the star formation rate density as a function of redshift, 
\begin{equation}\label{eq:sfr}
    \psi(z) = 10^{-2} \frac{(1+z)^{2.6}}{1 + \left[(1+z)/3.2\right]^{6.2}}\,\Msun\,\mathrm{yr}^{-1}\,\mathrm{Mpc}^{-3}. 
\end{equation}
Integrating over all formation redshifts up to $z_\mathrm{max}$, and converting to the number of mergers per unit time, gives us the local merger rate density 
\begin{equation}
    \mathcal{R}_{\mathrm{loc},\,i} = \frac{1}{t_\mathrm{l}(z_\mathrm{loc})} \int_{0}^{z_\mathrm{max}} \psi(z) f_{\mathrm{loc},\,i}(z) \frac{\mathrm{d}t_\mathrm{l}}{\mathrm{d} z} \mathrm{d}z = \frac{1}{H_0 t_\mathrm{l}(z_\mathrm{loc})} \int_{0}^{z_\mathrm{max}} \frac{\psi(z) f_{\mathrm{loc},\,i}(z)}{(1+z) E(z)} \mathrm{d}z,
\end{equation}
with $E(z) = \left[\Omega_\mathrm{rad}(1+z)^4 + \Omega_\mathrm{m}(1+z)^3 + \Omega_\mathrm{k}(1+z)^2 + \Omega_\mathrm{\Lambda} \right]^{1/2}$. 
In practice, we discretize this integral with $1000$ log-spaced redshift bins  between $z_\mathrm{loc}$ and $z_\mathrm{max}$. 
We consider systems out to $z_\mathrm{max} = 15$ because systems born at earlier times make up only $\approx\,1\%$ of the cumulative star formation and can thus be neglected. 

Each population is run at a single metallicity and allows all binaries to evolve for the entire age of the universe, allowing for rate calculations to be performed in post-processing. 
We simulate $16$ log-spaced metallicities between $Z_\odot/200$ and $2\,Z_\odot$ for each population model assumption. 
To account for metallicity evolution over cosmic time, we use the mean mass-weighted metallicity as a function of redshift in \citet{Madau2017a}, 
\begin{equation}\label{eq:met}
    \log_{10}\left\langle Z/Z_\odot \right\rangle  = 0.153 - 0.074 z^{1.34},
\end{equation}
and assume a truncated log-normal distribution of metallicities at each redshift with a dispersion of 0.5 dex~\citep{Bavera2020} that reflects over boundaries at $Z_\mathrm{min}=Z_\odot/200$ and $Z_\mathrm{max}=2\,Z_\odot$. 
The weights for each metallicity model $j$ at a given redshift $p(Z_j|z)$ (which are normalized to unity to account for our discrete metallicity models, $\sum_j p(Z_j|z) = 1$) 
are then folded into the rate calculation to give a local volumetric merger rate across all metallicity models, 
\begin{equation}
    \mathcal{R}_{\mathrm{loc},\,i} \simeq \frac{1}{H_0 t_\mathrm{l}(z_\mathrm{loc})} \sum_k \frac{\psi(\bar{z}_k) \sum_{j}p(Z_j | \bar{z}_k) f_{\mathrm{loc},\,i}(\bar{z}_k,Z_j)}{(1+\bar{z}_k) E(\bar{z}_k)} \Delta z_k,
\end{equation}
where $\bar{z}_k$ is the midpoint (in log space) of the $k$th redshift bin and $\Delta z_k$ is the size of the $k$th redshift bin. 

Local merger rates for each population model we simulate are shown in Table~\ref{tab:table}. 
In addition to \ac{BBH}, \ac{NSBH}, and \ac{BNS} rates for each model, we also give a Narrow rate for GW190814-like systems (defined as $0.06\,\leq\,q\,\leq\,0.16$, $20\,\leq\,M_\mathrm{tot}/\Msun\,\leq\,30$) and a Broad GW190814-like systems (defined as $q\,\leq\,0.2$, $M_\mathrm{tot}/\Msun\,\geq\,20$).

\begin{center}
\input{table}
\end{center}

\bibliography{library}{}
\bibliographystyle{aasjournal}

\end{document}

%% file: commands.tex
\def\Msun{\ensuremath{\mathit{M_\odot}}\xspace}

%% file: values.tex
\def\PrimaryMassApprox{\ensuremath{{\simeq\,23\,M_\odot}}\xspace}
\def\SecondaryMassApprox{\ensuremath{{\simeq\,2.6\,M_\odot}}\xspace}
\def\SecondaryMassNinety{\ensuremath{{2.50\text{--}2.67\,M_\odot}}\xspace}
\def\MassRatioApprox{\ensuremath{{\simeq\,0.11}}\xspace}
\def\MassRatioHigh{\ensuremath{{0.12}}\xspace}
\def\EventRedshift{\ensuremath{{\simeq\,0.053}}\xspace}

\def\MassRatioNinetyAprilGW{\ensuremath{{1.61\text{--}2.52\,M_\odot}}\xspace}

\def\CromatieMSP{\ensuremath{{2.05\text{--}2.24\,M_\odot}}\xspace}
\def\CromatieMSPUpdated{\ensuremath{{1.95\text{--}2.13\,M_\odot}}\xspace}
\def\FriereMSP{\ensuremath{{2.52\text{--}2.95\,M_\odot}}\xspace}

\def\DelayedqTenPercentApprox{\ensuremath{{\gtrsim\,0.4}}\xspace}
\def\EventOutlierStatus{\ensuremath{{99.9\%}}\xspace}
\def\FullPopMassRatiosLikeEvent{\ensuremath{{\approx\,10^{3}}}\xspace}
\def\NSBHMassRatiosLikeEvent{\ensuremath{{\approx\,8\%}}\xspace}
\def\NSBHMassRatiosLikeEventEfficientCE{\ensuremath{{\approx\,0.7\%}}\xspace}
\def\BBHMassRatiosLikeEvent{\ensuremath{{\lesssim\,0.03\%}}\xspace}
\def\FullPopMtwoLikeEventAlphaFive{\ensuremath{\text{ninth}}\xspace}
\def\FullPopMtwoLikeEventAlphaOne{\ensuremath{\text{first}}\xspace}

\def\DelayedAlphaOneBroadRate{\ensuremath{{0.08~\mathrm{Gpc}^{-3}\,\mathrm{yr}^{-1}}}\xspace}
\def\DelayedAlphaOneNarrowRate{\ensuremath{{0.04~\mathrm{Gpc}^{-3}\,\mathrm{yr}^{-1}}}\xspace}
\def\DelayedAlphaFiveBroadDrop{\ensuremath{{1.8}}\xspace}
\def\DelayedAlphaFiveNarrowDrop{\ensuremath{{2.4}}\xspace}

\def\ChannelAPrimary{\ensuremath{{>\,30\,M_\odot}}\xspace}
\def\ChannelASecondary{\ensuremath{{>\,12\,M_\odot}}\xspace}
\def\ChannelAEccentricity{\ensuremath{{\gtrsim\,0.9}}\xspace}
\def\ChannelBMassRatio{\ensuremath{{0.45\,\lesssim\,q\,\lesssim\,0.7}}\xspace}
\def\ChannelAFracLowMet{\ensuremath{{57\%}}\xspace}
\def\ChannelBFracLowMet{\ensuremath{{43\%}}\xspace}
\def\ChannelAFracHighMet{\ensuremath{{73\%}}\xspace}
\def\ChannelBFracHighMet{\ensuremath{{27\%}}\xspace}

%% file: acronyms.tex
\acrodef{GW}{gravitational-wave}
\acrodef{LIGO}{Laser Interferometer Gravitational-wave Observatory}
\acrodef{BH}{black hole}
\acrodef{BBH}{binary black hole}
\acrodef{NS}{neutron star}
\acrodef{BNS}{binary neutron star}
\acrodef{NSBH}{neutron star--black hole}
\acrodef{XRB}{X-ray binary}
\acrodef{SN}{supernova}
\acrodefplural{SN}[SNe]{supernovae}
\acrodef{PPI}{pulsational pair instability}
\acrodefplural{PPI}[PPIs]{pulsational pair instabilities}
\acrodef{O1}{first observing run}
\acrodef{O2}{second observing run}
\acrodef{O3}{third observing run}
\acrodef{O3a}{first half of the third observing run}
\acrodef{SNR}{signal-to-noise ratio}
\acrodef{GR}{general relativity}
\acrodef{PSD}{power spectral density}
\acrodef{CE}{common envelope}
\acrodef{EOS}{equation of state}
\acrodef{ZAMS}{zero-age main sequence}
\acrodef{LVC}{LIGO Scientific \& Virgo Collaboration}
\acrodef{NSF}{National Science Foundation}
\acrodef{LMG}{lower mass gap}

%% file: table.tex
\begin{deluxetable*}{@{\extracolsep{4pt}} c c c c c  c c c c c}
\label{tab:table}
\tablecaption{Population model assumptions and local merger rates across all simulated models. 
\ac{BBH}, \ac{NSBH}, and \ac{BNS} rates are highlighted in \textcolor{tablegreen}{green} if they are within the empirical $90\%$ credible bounds from LIGO--Virgo in \citet{O2RandP}, \citet{GWTC1}, and  \citet{GW190425}, respectively, are highlighted in \textcolor{tableorange}{orange} if they are a factor of $2$ above or below the $90\%$ credible bounds, and highlighted in \textcolor{tablered}{red} otherwise.
The following abbreviations are used for initial conditions sampling and natal kicks: MdS2017 for the multidimensional initial conditions from \citet{Moe2017}, S+2012 for the independent initial conditions from \citet{Sana2012}, and GM2020 for the kick prescription from \citet{Giacobbo2020}. 
\emph{Narrow} and \emph{Broad} rates for GW190814-like systems are described in Sec.~\ref{subsec:rate_predictions}. }
\tablehead{
\colhead{} & \multicolumn{3}{c}{Model Assumptions} & \colhead{} & \colhead{} & \multicolumn{3}{c}{Local Merger Rates [$\mathrm{Gpc^{-3}\,yr^{-1}}$]} & \colhead{} \\
\cline{1-5} \cline{6-10}
{Initial} & {\ac{CE}} & {\ac{CE}} & {Remnant} & {Natal} & {\ac{BBH}} & {\ac{NSBH}} & {\ac{BNS}} & {GW190814} & {GW190814} \\
{conditions} & {survival} & {efficiency} & {mass} & {kicks} & {$m_1 > 3\,\Msun$} & {$m_1 > 3\ \Msun$} & {$m_1 \leq 3\,\Msun$} & {\emph{Narrow}} & {\emph{Broad}} \\
{} & {} & {($\alpha$)} & {} & {} & {$m_2 > 3\,\Msun$} & {$m_2 \leq 3\,\Msun$} & {$m_2 \leq 3\,\Msun$} & {} & {} 
}
\startdata
MdS2017 & Optimistic & 1.0 & Delayed & GM2020 & \textcolor{tablered}{$2.0\,\times\,10^{3}$} & \textcolor{tablegreen}{$2.3\,\times\,10^{2}$} & \textcolor{tablegreen}{$1.4\,\times\,10^{3}$} & $2.4\,\times\,10^{-1}$ & $5.0\,\times\,10^{-1}$\\
    MdS2017 & Optimistic & 1.0 & Delayed & Bimodal & \textcolor{tablered}{$1.2\,\times\,10^{3}$} & \textcolor{tablegreen}{$2.1\,\times\,10^{1}$} & \textcolor{tablegreen}{$1.0\,\times\,10^{3}$} & $2.8\,\times\,10^{-1}$ & $9.1\,\times\,10^{-1}$\\
    MdS2017 & Optimistic & 1.0 & Rapid & GM2020 & \textcolor{tablered}{$2.3\,\times\,10^{3}$} & \textcolor{tablegreen}{$4.4\,\times\,10^{2}$} & \textcolor{tablegreen}{$1.4\,\times\,10^{3}$} & $6.0\,\times\,10^{-1}$ & $1.3\,\times\,10^{0}$\\
    MdS2017 & Optimistic & 1.0 & Rapid & Bimodal & \textcolor{tablered}{$1.6\,\times\,10^{3}$} & \textcolor{tablegreen}{$2.3\,\times\,10^{2}$} & \textcolor{tablegreen}{$1.0\,\times\,10^{3}$} & $3.1\,\times\,10^{-1}$ & $7.7\,\times\,10^{-1}$\\
    MdS2017 & Optimistic & 5.0 & Delayed & GM2020 & \textcolor{tablered}{$6.9\,\times\,10^{3}$} & \textcolor{tableorange}{$1.1\,\times\,10^{3}$} & \textcolor{tablered}{$8.9\,\times\,10^{3}$} & $4.8\,\times\,10^{0}$ & $6.8\,\times\,10^{0}$\\
    MdS2017 & Optimistic & 5.0 & Delayed & Bimodal & \textcolor{tablered}{$4.6\,\times\,10^{3}$} & \textcolor{tablegreen}{$2.0\,\times\,10^{2}$} & \textcolor{tableorange}{$4.6\,\times\,10^{3}$} & $3.1\,\times\,10^{0}$ & $7.4\,\times\,10^{0}$\\
    MdS2017 & Optimistic & 5.0 & Rapid & GM2020 & \textcolor{tablered}{$6.7\,\times\,10^{3}$} & \textcolor{tablegreen}{$4.2\,\times\,10^{2}$} & \textcolor{tablered}{$8.4\,\times\,10^{3}$} & $7.1\,\times\,10^{0}$ & $1.2\,\times\,10^{1}$\\
    MdS2017 & Optimistic & 5.0 & Rapid & Bimodal & \textcolor{tablered}{$5.7\,\times\,10^{3}$} & \textcolor{tablegreen}{$2.7\,\times\,10^{2}$} & \textcolor{tableorange}{$5.0\,\times\,10^{3}$} & $3.0\,\times\,10^{0}$ & $5.6\,\times\,10^{0}$\\
    MdS2017 & Pessimistic & 1.0 & Delayed & GM2020 & \textcolor{tablered}{$3.6\,\times\,10^{2}$} & \textcolor{tablegreen}{$4.4\,\times\,10^{1}$} & \textcolor{tablegreen}{$1.0\,\times\,10^{3}$} & $3.7\,\times\,10^{-2}$ & $6.4\,\times\,10^{-2}$\\
    MdS2017 & Pessimistic & 1.0 & Delayed & Bimodal & \textcolor{tablered}{$2.4\,\times\,10^{2}$} & \textcolor{tablegreen}{$5.5\,\times\,10^{0}$} & \textcolor{tablegreen}{$8.5\,\times\,10^{2}$} & $5.8\,\times\,10^{-2}$ & $1.2\,\times\,10^{-1}$\\
    MdS2017 & Pessimistic & 1.0 & Rapid & GM2020 & \textcolor{tablered}{$4.0\,\times\,10^{2}$} & \textcolor{tablegreen}{$5.5\,\times\,10^{1}$} & \textcolor{tablegreen}{$1.0\,\times\,10^{3}$} & $6.0\,\times\,10^{-2}$ & $1.5\,\times\,10^{-1}$\\
    MdS2017 & Pessimistic & 1.0 & Rapid & Bimodal & \textcolor{tablered}{$3.1\,\times\,10^{2}$} & \textcolor{tablegreen}{$2.6\,\times\,10^{1}$} & \textcolor{tablegreen}{$7.7\,\times\,10^{2}$} & $5.0\,\times\,10^{-2}$ & $1.2\,\times\,10^{-1}$\\
    MdS2017 & Pessimistic & 5.0 & Delayed & GM2020 & \textcolor{tablered}{$4.8\,\times\,10^{2}$} & \textcolor{tablegreen}{$1.4\,\times\,10^{2}$} & \textcolor{tableorange}{$2.8\,\times\,10^{3}$} & $9.1\,\times\,10^{-2}$ & $1.6\,\times\,10^{-1}$\\
    MdS2017 & Pessimistic & 5.0 & Delayed & Bimodal & \textcolor{tablered}{$2.8\,\times\,10^{2}$} & \textcolor{tablegreen}{$2.0\,\times\,10^{1}$} & \textcolor{tablegreen}{$1.7\,\times\,10^{3}$} & $5.1\,\times\,10^{-2}$ & $9.4\,\times\,10^{-2}$\\
    MdS2017 & Pessimistic & 5.0 & Rapid & GM2020 & \textcolor{tablered}{$4.4\,\times\,10^{2}$} & \textcolor{tablegreen}{$4.5\,\times\,10^{1}$} & \textcolor{tableorange}{$3.1\,\times\,10^{3}$} & $1.2\,\times\,10^{-1}$ & $2.0\,\times\,10^{-1}$\\
    MdS2017 & Pessimistic & 5.0 & Rapid & Bimodal & \textcolor{tablered}{$3.7\,\times\,10^{2}$} & \textcolor{tablegreen}{$2.5\,\times\,10^{1}$} & \textcolor{tablegreen}{$2.1\,\times\,10^{3}$} & $7.7\,\times\,10^{-2}$ & $1.0\,\times\,10^{-1}$\\
    S+2012 & Optimistic & 1.0 & Delayed & GM2020 & \textcolor{tablered}{$1.2\,\times\,10^{3}$} & \textcolor{tablegreen}{$1.3\,\times\,10^{2}$} & \textcolor{tablegreen}{$1.0\,\times\,10^{3}$} & $1.5\,\times\,10^{-1}$ & $2.4\,\times\,10^{-1}$\\
    S+2012 & Optimistic & 1.0 & Delayed & Bimodal & \textcolor{tablered}{$7.2\,\times\,10^{2}$} & \textcolor{tablegreen}{$1.1\,\times\,10^{1}$} & \textcolor{tablegreen}{$8.1\,\times\,10^{2}$} & $2.0\,\times\,10^{-1}$ & $6.2\,\times\,10^{-1}$\\
    S+2012 & Optimistic & 1.0 & Rapid & GM2020 & \textcolor{tablered}{$1.5\,\times\,10^{3}$} & \textcolor{tablegreen}{$1.7\,\times\,10^{2}$} & \textcolor{tablegreen}{$1.1\,\times\,10^{3}$} & $2.6\,\times\,10^{-1}$ & $5.5\,\times\,10^{-1}$\\
    S+2012 & Optimistic & 1.0 & Rapid & Bimodal & \textcolor{tablered}{$1.0\,\times\,10^{3}$} & \textcolor{tablegreen}{$7.0\,\times\,10^{1}$} & \textcolor{tablegreen}{$7.9\,\times\,10^{2}$} & $1.0\,\times\,10^{-1}$ & $3.6\,\times\,10^{-1}$\\
    S+2012 & Optimistic & 5.0 & Delayed & GM2020 & \textcolor{tablered}{$2.4\,\times\,10^{3}$} & \textcolor{tablegreen}{$5.1\,\times\,10^{2}$} & \textcolor{tableorange}{$5.0\,\times\,10^{3}$} & $9.8\,\times\,10^{-1}$ & $1.5\,\times\,10^{0}$\\
    S+2012 & Optimistic & 5.0 & Delayed & Bimodal & \textcolor{tablered}{$1.6\,\times\,10^{3}$} & \textcolor{tablegreen}{$7.2\,\times\,10^{1}$} & \textcolor{tableorange}{$3.1\,\times\,10^{3}$} & $5.8\,\times\,10^{-1}$ & $1.4\,\times\,10^{0}$\\
    S+2012 & Optimistic & 5.0 & Rapid & GM2020 & \textcolor{tablered}{$2.3\,\times\,10^{3}$} & \textcolor{tablegreen}{$1.8\,\times\,10^{2}$} & \textcolor{tableorange}{$5.1\,\times\,10^{3}$} & $1.3\,\times\,10^{0}$ & $2.2\,\times\,10^{0}$\\
    S+2012 & Optimistic & 5.0 & Rapid & Bimodal & \textcolor{tablered}{$2.0\,\times\,10^{3}$} & \textcolor{tablegreen}{$7.7\,\times\,10^{1}$} & \textcolor{tableorange}{$3.4\,\times\,10^{3}$} & $4.7\,\times\,10^{-1}$ & $1.0\,\times\,10^{0}$\\
    S+2012 & Pessimistic & 1.0 & Delayed & GM2020 & \textcolor{tableorange}{$1.7\,\times\,10^{2}$} & \textcolor{tablegreen}{$3.3\,\times\,10^{1}$} & \textcolor{tablegreen}{$7.9\,\times\,10^{2}$} & $4.9\,\times\,10^{-2}$ & $7.7\,\times\,10^{-2}$\\
    S+2012 & Pessimistic & 1.0 & Delayed & Bimodal & \textcolor{tablegreen}{$8.4\,\times\,10^{1}$} & \textcolor{tablegreen}{$3.7\,\times\,10^{0}$} & \textcolor{tablegreen}{$6.3\,\times\,10^{2}$} & $4.1\,\times\,10^{-2}$ & $8.3\,\times\,10^{-2}$\\
    S+2012 & Pessimistic & 1.0 & Rapid & GM2020 & \textcolor{tableorange}{$2.0\,\times\,10^{2}$} & \textcolor{tablegreen}{$3.3\,\times\,10^{1}$} & \textcolor{tablegreen}{$8.7\,\times\,10^{2}$} & $4.6\,\times\,10^{-2}$ & $9.7\,\times\,10^{-2}$\\
    S+2012 & Pessimistic & 1.0 & Rapid & Bimodal & \textcolor{tableorange}{$1.3\,\times\,10^{2}$} & \textcolor{tablegreen}{$1.1\,\times\,10^{1}$} & \textcolor{tablegreen}{$6.0\,\times\,10^{2}$} & $2.3\,\times\,10^{-2}$ & $7.7\,\times\,10^{-2}$\\
    S+2012 & Pessimistic & 5.0 & Delayed & GM2020 & \textcolor{tableorange}{$1.7\,\times\,10^{2}$} & \textcolor{tablegreen}{$9.4\,\times\,10^{1}$} & \textcolor{tablegreen}{$1.7\,\times\,10^{3}$} & $3.8\,\times\,10^{-2}$ & $5.9\,\times\,10^{-2}$\\
    S+2012 & Pessimistic & 5.0 & Delayed & Bimodal & \textcolor{tablegreen}{$8.6\,\times\,10^{1}$} & \textcolor{tablegreen}{$1.1\,\times\,10^{1}$} & \textcolor{tablegreen}{$1.3\,\times\,10^{3}$} & $1.7\,\times\,10^{-2}$ & $4.7\,\times\,10^{-2}$\\
    S+2012 & Pessimistic & 5.0 & Rapid & GM2020 & \textcolor{tableorange}{$1.5\,\times\,10^{2}$} & \textcolor{tablegreen}{$3.1\,\times\,10^{1}$} & \textcolor{tablegreen}{$2.0\,\times\,10^{3}$} & $3.9\,\times\,10^{-2}$ & $6.3\,\times\,10^{-2}$\\
    S+2012 & Pessimistic & 5.0 & Rapid & Bimodal & \textcolor{tableorange}{$1.3\,\times\,10^{2}$} & \textcolor{tablegreen}{$1.3\,\times\,10^{1}$} & \textcolor{tablegreen}{$1.5\,\times\,10^{3}$} & $3.9\,\times\,10^{-2}$ & $6.7\,\times\,10^{-2}$
    \enddata
\end{deluxetable*}